\documentclass[twocolumn,superscriptaddress,amsmath,showpacs,amssymb]{revtex4}

\usepackage{graphicx}
\usepackage{epsfig}
\usepackage{dcolumn}

\begin{document}

\title{A diffusion Monte Carlo study of small para-Hydrogen clusters}
\date{\today}

\author{Rafael Guardiola}
\affiliation{ Departamento de F\'{\i}sica At\'omica y Nuclear, Facultad de
F\'{\i}sica, 46100 Burjassot, Spain}

\author{Jes\'us Navarro}
\affiliation{ IFIC (CSIC-Universidad de Valencia), Apdo. 22085,
46071 Valencia, Spain}

\pacs{67.40.Db, 36.40.-c, 61.46.Bc}

\begin{abstract}
Ground state energies and chemical potentials of parahydrogen clusters are
calculated from 3 to 40 molecules using the diffusion Monte Carlo technique
with two different $p$-H$_2$--$p$-H$_2$ interactions.
This calculation improves a previous one by the inclusion of three-body correlations
in the importance sampling, by the time step adjustement and by a better estimation 
of the statistical errors. Apart from the cluster with 13 molecules, 
no other magic clusters are predicted, in contrast with path integral Monte 
Carlo results. 
\end{abstract}

\maketitle

Theoretical studies of parahydrogen clusters have attracted a 
growing interest in the past years, partly motivated by a recent 
experiment~\cite{tejeda2004} in which Raman scattering was used in cryogenic
free jets of the pure gas. Small changes in the frequency
near the $Q_1(0)$ line of the monomer, were observed and 
interpreted as intermolecular effects on the intramolecular potential. 
($p$-H$_2$)$_N$ clusters with $N=2-8$ were clearly identified 
through frequency shifts ranging from $\Delta \nu = -0.40 {\rm cm}^{-1}$ 
for $N=2$ to $\Delta \nu = -2.35 {\rm cm}^{-1}$ for $N=8$.   
The experiment also showed a bump at $N=13$, $N=33$ 
and $N=55$, which were interpreted as a signal of magical clusters. 
However, it is worth mentioning that these three values are actually 
extrapolations from smaller clusters and presumably are approximate.

Magical numbers appear in classical Lennard-Jones clusters, related
to geometrical shapes~\cite{baletto2005}. Several papers have appeared 
in the last year with the main objective of checking the magical numbers 
found in Ref.~\cite{tejeda2004}, and/or studying possible superfluidity effects
in parahydrogen clusters. Indeed, Path Integral Monte Carlo (PIMC) 
calculations~\cite{sindzingre1991} have found a large superfluid fraction
in clusters with N=13 and 18 at temperatures $T\leq2$~K.  
A superfluid response has been observed in small clusters consisting
of a carbonyl sulfide cromophore surrounded by 15-17 $p$-H$_2$ molecules,
all within a large helium droplet~\cite{grebenev2000}. This has been confirmed by 
several MC simulations~\cite{kwon2002,paesani2003,tang2004,paesani2005,baroni2005}
of doped $p$-H$_2$ clusters.

Systematic studies of 
($p$-H$_2$)$_N$ clusters, covering the range from $N=3$ 
to $N=50$ molecules, have been done based on powerful many-body techniques,
as difusion Monte Carlo (DMC)~\cite{guardiola2006},  
PIMC~\cite{mezzacapo2006,mezzacapo2007,khairallah2007}, and
PIMC adapted to the ground state (PIGS)~\cite{cuervo2006}. 
Whereas up to $N\simeq22$ all these calculations are substantially in agreement,
for heavier clusters there are noticeable differences between DMC and PIMC
results, particularly for $N \geq 26$. PIMC chemical potentials show very
prominent peaks at N=26, 29, 34 and 39, in contrast with the smooth
behavior obtained with DMC. 

In this work we present new DMC calculations, improving our 
previous ones~\cite{guardiola2006} so as to get very precise 
results within our computational capacity. Specifically, we consider
three aspects: the importance sampling function, 
the time step adjustement and the estimation of the statistical errors. 

The DMC procedure is significantly improved when using a good 
importance sampling wave function, the main effect being the reduction 
of the variance of the stochastic procedure. We have used a Jastrow function with 
two- and three-body correlations:
\begin{equation}
\Phi_T = \exp( \bar{ u}_2 + u_3) ,
\label{JWF}
\end{equation}
where
\begin{equation}
\bar{ u}_2 =\sum_{i<j} 
\left[u_2(r_{ij}) + \lambda_T \xi^2(r_{ij}) r_{ij}^2\right] ,
\end{equation}
with
\begin{equation}
u_2(r) = - \sum_{i<j}\left[\frac{p_5}{r_{ij}^5} + \frac{p_1}{ r_{ij}}
\right] , 
\end{equation}
\begin{equation}
\xi(r) =
\exp \left( - \frac {(r-s_T)^2}{w_T^2} \right) ,
\end{equation}
and
\begin{equation}
u_3 = - \frac{\lambda_T}{2} \sum_\ell {\bf G}_\ell {\bf G}_\ell ,
\end{equation}
with
\begin{equation}
{\bf G}_\ell = \sum_{i\neq \ell} \xi(r_{li}) {\bf r}_{li} .
\end{equation}
Indices $i, j, l$ run over the number of molecules in the cluster. 
This function is described in terms of five variational parameters, 
$p_5, p_1, s_T, \omega_T$, and $\lambda_T$. In our previous 
calculations~\cite{guardiola2006} we used
the standard two-body Jastrow function $\Phi_T = \exp(u_2)$. The present
trial function includes an enlarged the two-body variational space
and also three-body correlations in the form suggested 
in Ref.~\cite{schmidt1981}, which still has ${\cal O}(N^2)$ computational 
complexity. 

DMC is based in a short-time approximation of the Green's function related 
to the imaginary time Schr\"odinger equation. In this way, an initial wave 
function $\Phi_T(t=0)$ evolves to the exact ground state wave function 
$\Psi$ at large $t$ after many short-time steps $\tau$. 
We have used the ${\cal O}(\tau^3)$ approximation to the Green's function
as described in Refs.~\cite{vrbik1986,chin1990}, which provides energies 
${\cal O}(\tau^2)$. 
The time step adjustment is the following: from calculations
at the relative large steps $0.001$ and $0.0005 {\rm K}^{-1}$,
we obtain the Richardson extrapolated value
\begin{equation}
\frac{1}{3} \left( 4 E(0.0005) - E(0.001) \right) \, , 
\end{equation}
based on the $\tau$ expansion $E(\tau) = E(0) + C \tau^2 + \cdots$.
This value turns out to be very close to the calculations with much 
smaller time steps, as it may be checked in the last three rows 
of Table~\ref{timestep}. This checks that the algorithm behaves 
as ${\cal O}(\tau^2)$, as expected, and suggests to use the value 
$\tau=0.0001 {\rm K}^{-1}$ for massive calculations with a negligible bias.

\begin{table}[h!] 
\caption{
\label{timestep}
Determination of optimal time step $\tau$ from different evaluations of 
the binding energy B(N) of several clusters. The row labelled 
{\em R.E.} is the Richardson extrapolated value obtained from the 
previous two rows. 
The statistical standard deviation is indicated in
parenthesis (error in the final digit shown).
Energies and statistical errors are in K.}
\begin{ruledtabular}
{\begin{tabular}{llll}
$\tau$ &  $ B(10)$   &    $ B(20)$  &   $ B(30)$ \\
0.001    &183.47(5)  &   559.28(17) &    1006.4(3) \\
0.0005   &185.91(6)  &   566.56(17) &    1020.0(4) \\
{\em R.E.} &186.72(9) &   568.99(28) &    1024.5(5) \\
0.0001   &186.93(6)  &   569.16(12) &    1025.2(2) \\
0.00002  &186.72(3)  &   569.48(7)  &    1024.8(1) \\
\end{tabular}}
\end{ruledtabular}
\end{table}

A further improvement of the calculation regards the estimate of the 
statistical error. Because of the sequential Markov chain nature of Monte Carlo 
algorithms, successive samples are strongly correlated, and the typical 
way of estimating the variance, 
$\sigma^2 = \langle H^2\rangle - \langle H\rangle^2$, 
may be too optimistic.
To avoid these correlations we computed a number of times (10, typically),
the binding energies, with independent and randomized runs, and estimate 
the variance from these results. This requires a considerable increasing in 
computational time, but the obtained standard deviations are very precisely 
computed. Specifically, we have used 1000 walkers with 10$^5$ steps 
plus 20000 stabilization steps in each walker.

Hydrogen molecules interact through weak van der Waals forces that, 
nevertheless, are sufficiently strong to bound clusters with any number 
of molecules. Several forms have been derived to describe the $p$-H$_2$--$p$-H$_2$ 
interaction. Two of them are of particular interest because they combine 
{\em ab initio} properties with properties of the gas (or solid) as well 
as experimental information from collisions, one due to Silvera and
Goldman~\cite{silvera1978}, and the other to Buck {\em et. al.}~\cite{buck1983},
hereafter referred to as SG and BHKOS, respectively. The main difference 
among them is that the former contains a repulsive long-range term ($c_9/r^9$) 
with the objective of providing  an approximation for the effective 
potential in a solid. Recent calculations have employed 
BHKOS potential~\cite{guardiola2006}, SG
potential~\cite{mezzacapo2006,mezzacapo2007,khairallah2007} 
or both~\cite{cuervo2006}. We present here results with both interactions.

\begin{table}[h!]
\caption{\label{table1}
DMC ground state (p-H$_2$)$_N$ binding energies (in K) obtained 
with BHKOS interaction~\cite{buck1983}.} 
\begin{ruledtabular}
\begin{tabular}{dddddd}
$N$&           &$N$ &            &$N$ &  \\ \hline
 2 &  4.3114   & 15 & 371.17(3)  & 28 &  931.46(18) \\
 3 &  14.66(1) & 16 & 408.56(5)  & 29 &  978.50(10) \\
 4 &  30.50(1) & 17 & 446.68(3)  & 30 & 1025.66(14) \\
 5 &  50.33(1) & 18 & 486.34(8)  & 31 & 1074.03(30) \\
 6 &  73.40(1) & 19 & 527.46(8)  & 32 & 1122.19(23) \\
 7 &  98.76(1) & 20 & 569.72(6)  & 33 & 1170.42(38) \\
 8 & 126.16(1) & 21 & 612.27(11) & 34 & 1219.26(25) \\
 9 & 155.51(2) & 22 & 655.65(12) & 35 & 1267.74(35) \\
10 & 186.86(3) & 23 & 700.50(13) & 36 & 1317.87(15) \\
11 & 220.98(2) & 24 & 745.63(14) & 37 & 1366.26(42) \\
12 & 257.94(2) & 25 & 791.57(16) & 38 & 1416.12(22) \\
13 & 297.80(9) & 26 & 837.88(29) & 39 & 1465.55(42) \\
14 & 334.24(5) & 27 & 884.71(15) & 40 & 1516.00(37) \\
\end{tabular}
\end{ruledtabular}
\end{table}

\begin{table}[h!]
\caption{\label{table2}
DMC ground state (p-H$_2$)$_N$ binding energies (in K) obtained 
with the SG interaction~\cite{silvera1978}.} 
\begin{ruledtabular}
\begin{tabular}{dddddd}
$N$&            & $N$&             & $N$&            \\ \hline
 2 &   3.8456   & 15 &  341.12(5)  & 28 &  860.69(12) \\
 3 &   13.22(1) & 16 &  375.81(6)  & 29 &  904.56(10) \\
 4 &   27.59(1) & 17 &  411.56(4)  & 30 &  949.06(21) \\
 5 &   45.73(1) & 18 &  448.24(5)  & 31 &  993.46(12) \\
 6 &   66.87(1) & 19 &  486.31(5)  & 32 & 1038.04(25) \\
 7 &   90.14(1) & 20 &  525.35(7)  & 33 & 1083.36(22) \\
 8 &  115.42(1) & 21 &  564.92(10) & 34 & 1128.38(32) \\
 9 &  142.46(2) & 22 &  605.40(8)  & 35 & 1174.29(23) \\
10 &  171.41(3) & 23 &  646.60(8)  & 36 & 1220.32(22) \\
11 &  202.74(1) & 24 &  688.50(11) & 37 & 1266.42(20) \\
12 &  236.56(4) & 25 &  731.02(11) & 38 & 1312.22(21) \\
13 &  272.53(5) & 26 &  774.08(7)  & 39 & 1358.46(29) \\
14 &  306.63(4) & 27 &  817.34(14) & 40 & 1405.45(27) \\
\end{tabular}
\end{ruledtabular}
\end{table}

The calculated DMC ground state binding energies are displayed in 
Tables~\ref{table1} and~\ref{table2}, for BHKOS and SG potentials, 
respectively. As usual, the numbers in parenthesis are the errors in the 
final digit shown, and correspond to the standard deviation.
The binding energies for the dimer have been obtained by numerical 
integration of the Schr\"odinger equation. 

\begin{figure}[h!]
\caption{
(Color online)
Comparison of the binding energies per molecule obtained with
diffusion Monte Carlo (DMC) and a variational Monte Carlo (VMC)
calculation based on Eq.~(\ref{JWF}) with two- and three-body
correlations. The interaction is BHKOS. 
The error bars are within the symbol size.}
\label{VMC}
\centerline{
\includegraphics[width=9.5cm]{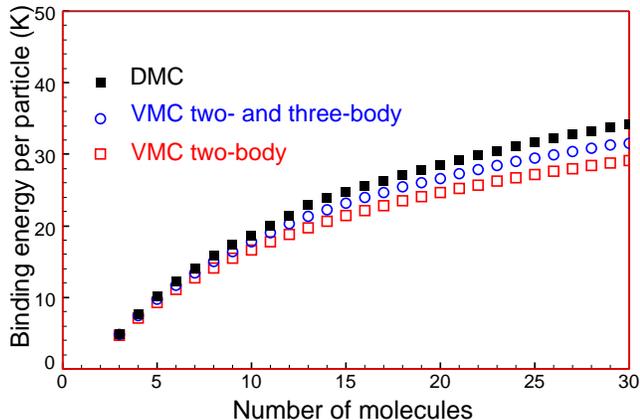}
}
\end{figure}

The inclusion of triplet correlations in the importance sampling function
leads to a noticeably improvement of the variational energies, as showed
in Fig.~\ref{VMC} for BHKOS potential. The DMC energies are basically
the same as those of Ref.~\cite{guardiola2006}, with slightly more binding 
in the heavier clusters~\cite{note}. The main difference lies in the reduction of
the standard deviation by typically a factor of 2.  

\begin{figure}[h!]
\caption{
(Color online)
DMC chemical potential (in K) of ($p$-H$_2$)$_N$ clusters as a function of 
the number $N$ of molecules. The error bars are within the 
symbol size.}
\label{mu}
\centerline{
\includegraphics[width=9.5cm]{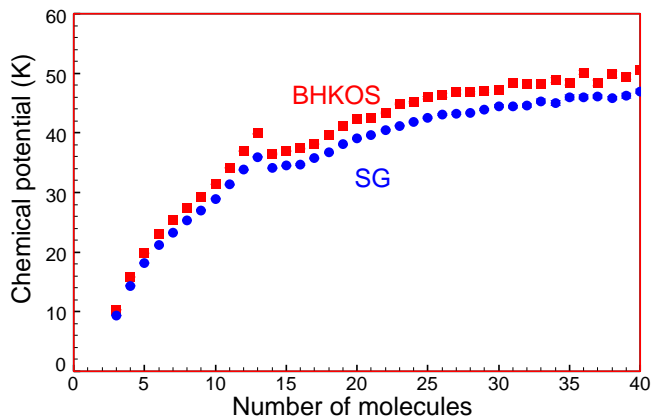}
}
\end{figure}

The total binding energies grow monotonically with the number of constituents.
In order to determine an enhanced stability related to magic sizes it is 
convenient to analyze the variation with $N$ of the dissociation energy or 
chemical potential, defined from the ground state energies $E(N)$ as
\begin{equation}
\label{chempotential}
\mu_N = E(N-1)-E(N)~.
\end{equation}
This quantity is plotted in Fig.~\ref{mu} as a function of the number of
molecules $N$. 
The main physical result of this figure is the presence of a neat peak
at $N=13$, indicating the magical character of this cluster. 
Although the two used interactions give different total energies,
the BHKOS potential providing more binding than the SG one,  
this peak is present for both interactions.

Beyond $N=13$ our calculations do not show any clear signal of local 
enhancement of the chemical potential. It should be mentioned that even 
if the relative error of the total energies is around 10$^{-4}$, the
relative error in the separation energies can be as high as 10$^{-2}$,
as a consequence of the strong cancelations appearing when computing $\mu$.
So, even after our formidable numerical effort, the absolute error
of $\mu$ for $N \simeq 40$ may be as high as 0.5~K. Having this
fact in mind, the only possible structure, apart from N=13, is N=36 for
BHKOS potential. However, $\mu_{36}$ is $\simeq 1.70\pm0.6$~K higher than
its neighbors $\mu_{35,37}$ and one cannot exclude that it could
simply be a statistical fluctuation. Consequently
the only magical cluster firmly established here is N=13, independent
of the interaction.

\begin{figure}[h!]
\caption{
(Color online)
DMC and PIMC chemical potentials of ($p$-H$_2$)$_N$ clusters 
as a function of the number $N$ of molecules, calculated with
SG interaction~\cite{silvera1978}. Filled circles and squares are
PIMC results of Ref.~\cite{khairallah2007}, the solid line corresponds
to PIMC results of Ref.~\cite{mezzacapo2007}. 
Error bars have not been drawn for the clarity of presentation.}
\label{compara}
\centerline{
\includegraphics[width=9.5cm]{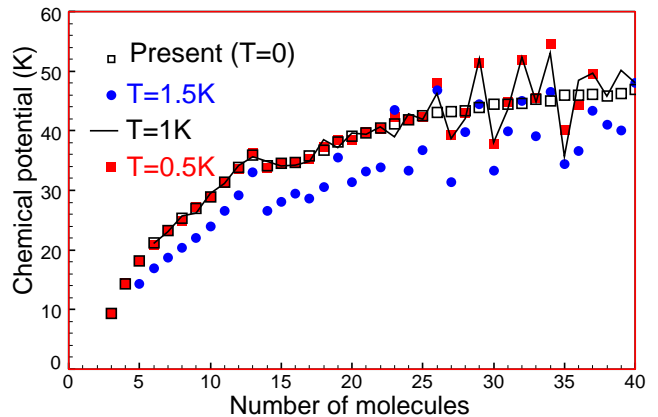}
}
\end{figure}

In contrast, PIMC calculations show very prominent 
peaks at $N=26,\ 29,\ 32,\ 34$ and 39, as show in Fig.~\ref{compara}. 
The PIMC results at $T=0.5$~K and 1.5~K are from Ref.~\cite{khairallah2007},
and those at $T=1$~K from Ref.~\cite{mezzacapo2007}. 
Up to $N\approx 25$, our DMC results are indistinguishable from the
the PIMC ones of Ref.~\onlinecite{khairallah2007} at $T=0.5$~K
or those of Ref.~\onlinecite{mezzacapo2007} at $T=1$~K. It is
worth noticing that these PIMC results at $T=0.5$ and 1~K are 
essentially identical in the calculated range of cluster sizes,
with the noticeable exceptions of $N=23$, 35 and 36. However, it
should be kept in mind that PIMC error bars have not been drawn
in Fig.~\ref{compara} for the sake of clarity. 

The existence of peaks in the chemical potential seems to be related to
thermal effects. These could manifest in enhanced stability thresholds
at finite temperature, similarly to what has been observed in 
$^4$He droplets~\cite{bruhl2004}. But according to Ref.~\cite{mezzacapo2007} 
such thermal effects should be associated to a 
coexistence of solid-like and liquid-like phases, with a dominance
of the latter at low T, as a result of both the zero-point 
motion and quantum permutation exchanges. To this respect it is worth recalling
that while DMC may be affected by the constraint imposed by the importance
sampling function, PIMC has no such constraint. Actually, our importance sampling
function is of the type used to describe liquid-like clusters. We have checked 
that the DMC ground state energy for the magical ($p$-H$_2$)$_{13}$ cluster 
does not change employing instead an importance sampling function where 
the molecules are localized at the vertex of the corresponding truncated 
MacKay icosahedra. This cluster is definitely liquid-like, in agreement with PIMC
results. It would be interesting to perform a similar solid-like DMC calculation
for $N \geq 26$. Although very computationally demanding, such a calculation could 
be useful to ascertain the phase of $p$-H$_2$ clusters in this size region.

The authors acknowledge stimulating conversations with J.P. Toennies and S. Montero
and correspondence with J.E. Cuervo.
This work is supported by grants FIS2004-00912 (MCyT/FEDER, Spain), and
ACOMP07-003 (Generalitat Valenciana, Spain).


\begin{thebibliography}{99}
\bibitem{tejeda2004}
G. Tejeda, J.M. Fern\'andez, S. Montero, D. Blume, and J. P. Toennies, 
Phys. Rev. Lett. {\bf 92}, 223401 (2004).
\bibitem{baletto2005}
F. Baletto and R. Ferrando,
Rev. Mod. Phys. {\bf 77}, 319 (2005).
\bibitem{sindzingre1991}
P. Sindzingre, D.M.Ceperley and M.L.Klein, 
Phys. Rev. Lett. {\bf  67}, 1871 (1991).
\bibitem{grebenev2000}
S. Grebenev, B. Sartakov, J.P. Toennies, and A.F. Vilesov,
Science {\bf 289}, 1532 (2000).
\bibitem{kwon2002}
Y. Kwon and K.B. Whaley,
Phys. Rev. Lett. {\bf 89}, 273401 (2002).
\bibitem{paesani2003}
F. Paesani, R.E. Zillich and K.B. Whaley,
J. Chem. Phys. {\bf 119}, 11682 (2003).
\bibitem{tang2004}
J. Tang and A.R.W. McKellar,
J. Chem. Phys. {\bf 121}, 3087 (2004).
\bibitem{paesani2005}
F. Paesani, R.E. Zillich, Y. Kwon, and K.B. Whaley,
J. Chem. Phys. {\bf 122}, 181106 (2005).
\bibitem{baroni2005}
S. Baroni and S. Moroni,
ChemPhysChem {\bf 6}, 1884 (2005].
\bibitem{guardiola2006}
R. Guardiola and J. Navarro,
Phys. Rev. A {\bf 74}, 025201 (2006). 
\bibitem{cuervo2006}
J.E. Cuervo and P.N. Roy, 
J. Chem. Phys. {\bf  125}, 124314 (2006).
\bibitem{mezzacapo2006}
F. Mezzacapo and M. Boninsegni, 
Phys. Rev. Lett. {\bf  97}, 045301 (2006).
\bibitem{mezzacapo2007}
F. Mezzacapo and M. Boninsegni, 
Phys. Rev. A  {\bf 75}, 033201 (2007).
\bibitem{khairallah2007}
S.A. Khairallah, M.B. Sevryuk, D.M.Ceperley, and J. P. Toennies,
Phys. Rev. Lett. {\bf 98}, 183401 (2007).
\bibitem{silvera1978}
I.F. Silvera, V.V. Goldman, 
J. Chem. Phys. {\bf  69}, 4209 (1978)
\bibitem{buck1983}
U. Buck, F. Huisken, A. Kohlhase, D. Otten, and J. Schaeffer,
J. Chem. Phys. {\bf  78}, 4439 (1983)
\bibitem{vrbik1986}
J. Vrbik and S.M. Rothstein, 
J. Comput. Phys. {\bf   63}, 130 (1986).
\bibitem{chin1990}
S.A. Chin, 
Phys. Rev. A {\bf   42}, 6991 (1990).
\bibitem{schmidt1981} 
K.E. Schmidt, M.A. Lee, M.H. Kalos, and G.V. Chester, 
Phys. Rev. Lett. {\bf 47}, 807 (1981).
\bibitem{note}
There is a misprint in Table I of Ref.~\cite{guardiola2006}. The entry for $B/N$ at 
$N=23$ should read 30.43(2), instead of the quoted 29.94(2). We acknowledge 
J.E. Cuervo for having detected it. See also Ref.~\cite{mezzacapo2007}.  
\bibitem{bruhl2004}
R. Br\"uhl, R. Guardiola, A. Kalinin, O. Kornilov, J. Navarro, 
T. Savas, and J.P. Toennies,
Phys. Rev. Lett. {\bf 92}, 185301 (2004).
\end{thebibliography}
\end{document}